\documentstyle[preprint,aps]{revtex}
\input epsf.tex
\def\DESepsf(#1 width #2){\epsfxsize=#2 \epsfbox{#1}}
\begin{document}
\preprint{\vbox{\hbox{OITS-604, UH-511-851-96}}}
\draft
\title{Single Isospin Decay Amplitude and CP Violation}
\author{$^1$N.G. Deshpande, $^2$Xiao-Gang He, and $^3$Sandip Pakvasa}
\address{$^1$Institute of Theoretical Science, University of Oregon\\
Eugene, OR 97403-5203, USA\\
$^2$School of Physics, University of Melbourne\\
Parkville, Vic. 3052, Australia\\
and\\
$^3$Department of Physics and Astronomy, University of Hawaii\\
Honolulu, HI 90822}
\date{June 1996}
\maketitle
\begin{abstract}
We study partial rate asymmetry in decays with single isospin final state.
For K meson or hyperon decays,
 the partial rate asymmetries are always zero if the final states are 
single isospin states. In B decays the situation 
is dramatically different and partial rate asymmetries can be non-zero even 
if the final states are single isospin states. We calculated partial rate 
asymmetries for several B decays with single isospin amplitude in the final 
states using factorization approximation. We find that in some cases the 
asymmetries can be large.
\end{abstract}
CP violation is one of the few remaining unresolved mysteries in 
particle physics. The explanation in the Standard Model (SM) based on 
Cabibbo-Kobayashi-Maskawa (CKM) matrix\cite{ckm} is still not established,
although there is no conflict between the observation of CP violation 
in the neutral K-system and theory\cite{1}. It is important to carry out more 
experiments to test CP violation in the SM. The study 
of CP violation in the $B$ system is very important which may provide crucial
information about CP violation\cite{stone}. The $B$ system offers 
several final states that provide a rich source for the study of this 
phenomena. In many cases CP violation in $B$ decays occurs in a quite 
different form from $K$ or hyperon
 decays. 
In this paper we study CP violating partial rate asymmetries in B decays.
 We clarify some subtleties for CP violation in partial 
rate asymmetry in relation to isospin analysis. 

The decay amplitudes for Kaon or hyperon are 
customarilly
parametrized according to isospin decomposition in the final states because 
isospin states are eigenstates of strong interaction. For 
example the amplitude for $K^0 \rightarrow \pi^+\pi^-$ decay can be 
parametrized as
\begin{eqnarray}
A = A_0e^{i\delta_w^0 +\delta_s^0} +  A_2e^{i\delta_w^2 +\delta_s^2}\;,
\label{iso}
\end{eqnarray}
where superscript $0$ and $2$ indicate the isospin of the final states. 
$\delta_w^i$ and $\delta_s^i$ are the CP violating weak and CP conserving 
strong phases, respectively.

The partial rate asymmetry $A_{asy}$ is given by
\begin{eqnarray}
A_{asy} = {|A|^2 - |\bar A|^2\over |A|^2+|\bar A|^2}
 = -{2A_oA_2 \mbox{sin}(\delta_w^0-\delta_w^2)\mbox{sin}(\delta_s^0-\delta_s^2)
\over A_0^2+A_2^2 +2A_0A_2  \mbox{sin}(\delta_w^0-\delta_w^2)\mbox{sin}(\delta_s^0-\delta_s^2)}\;.
\end{eqnarray}
The conditions for non-zero partial rate 
asymmetry in this case are: there must exist at least
two different isospin decay amplitudes with different
weak, and strong 
rescattering phases. It is clear that for final states with single isospin,
for example $K^- \rightarrow \pi^- \pi^0$, the partial rate asymmetry vanishes.
 The same equation could be used for $B\rightarrow \pi\pi$ decays. 
But the argument is now incorrect. We wish to consider the difference 
in some detail.

We now consider the same process as above from the quark level.
In the SM the effective Hamiltonian responsible 
for $K\rightarrow \pi^+\pi^-$ at the quark level can be 
parametrized as\cite{buras1},
\begin{eqnarray}
H_{eff} = \sum_i[V_{us}V_{ud}^* c^u_i +  V_{cs}V_{cd}^*c^c_i +  
V_{ts}V_{td}^*c^t_i]O_i\;,
\end{eqnarray}
where $c_i^f$ are Wilson Coefficients (WC) of the corresponding quark 
operators $O_i$. At the one loop level $c_i^u$ contain the tree and u internal 
quark contributions,
$c_i^{c,t}$ contain internal c- and t- quark contributions. Since the $u$, 
$\bar u$
pair is lighter than s-quark, at the one loop level with u quark in the loop, 
absorptive 
amplitudes will be generated, and $c_i^u$ has a strong rescattering phase\cite{soni}. 
On the other hand
no absorptive parts exist for $c_i^{c,t}$. Naively, one would 
obtain partial rate asymmetry for $K^0\rightarrow \pi^+\pi^-$ which seems to have 
nothing to do with strong phase shifts in different isospin amplitudes.
However, this argument turns out to be wrong. 
It has been 
pointed out by Gerard and Hou\cite{gh} that CPT 
theorem is violated if one is not careful to include all diagrams of the same 
order\cite{other}. The interference term responsible for rate asymmetry due to
$c^u_i$ is an interference between penguin amplitudes of order 
$\alpha_s^2$. To this order there is also another contribution which is 
the interference between the 
tree amplitude and the higher order 
penguin amplitudes with absorptive part developed in the
vacuum polarization of the virtual gluon. This contribution cancels the previous
contribution.
In practical calculation, $c^u_i$ must be treated as real
in this case. The rule is that for the phase of any one of the penguin WC, if there
is a tree amplitude with the same CKM factor, the phase in that penguin WC must
be removed when the final states are the same as the tree amplitude.
In a more general formulation of the problem from CPT theorem and unitarity 
considerations, Wolfenstein showed that any diagonal strong phase (the phase due to 
rescattering of the states which are the same as the final states) do not contribute to
partial rate asymmetry\cite{wolf}. The phases in $c_i^u$ are diaganal phases 
in the present case\cite{wolf}.

At the hadron level, the decay amplitude is given by
\begin{eqnarray}
A(K^0\rightarrow \pi^+\pi^-) = \sum_i<\pi^+\pi^-|[V_{ud}^*V_{us} c^u_i +  
V_{cd}^*V_{cs} c^c_i +  V_{td}^*V_{ts} c^t_i]O_i|K>\;.
\end{eqnarray}
The strong rescattering phases
are generated by rescattering the two pions in the final states.
This is because in this case, only two pion
final states are opened in this kinematic region with the right parity. Three
pions in the intermediate states are allowed kinematically, but three
pions would not rescatter into two pions because of G parity conservation. 
The part responsible for absorptive amplitude is given by
\begin{eqnarray}
A(K^0\rightarrow \pi^+\pi^-) &=& \sum_i\sum_I<\pi^+\pi^-|[V_{ud}^*V_{us} c^u_i +  
V_{cd}^*V_{cs} c^c_i +  V_{td}^*V_{ts} c^t_i]|(\pi\pi)_I>\nonumber\\
&\times&<(\pi\pi)_I|O_i|K>\;,
\end{eqnarray}
where I is summed over isospin eigenstates.

Since isospin symmetry is respected by strong interaction, we can generally parametrize the 
hadronic matrix elements as, after the rescattering strong phase shifts are included,
\begin{eqnarray}
 &\sum_I&<\pi^+\pi^-|[V_{ud}^*V_{us} c^u_i +  
V_{cd}^*V_{cs} c^c_i +  V_{td}^*V_{ts} c^t_i]|(\pi\pi)_I><(\pi\pi)_I|O_i|K>
\nonumber\\ &=& [V_{ud}^*V_{us} c^u_i +  
V_{cd}^*V_{cs} c^c_i +  V_{td}^*V_{ts} c^t_i]
[x^0_ia_0e^{i\delta^0_s}+x^2_i a_2e^{i\delta^2_s}]\;,
\end{eqnarray}
where $a_{0,2}$ and $\delta_s^{0,2}$ are the isospin eigen-amplitudes and 
strong rescattering phases from all contributions, and
$x_i^{0,2}$ are the Clebsch-Gordan coefficients for each operator.
Knowing that the absorptive parts in $c_i^u$ are canceled by the
other effects, 
we should take all $c_i^{u,c,t}$ in the above equation to be real.
We have 
for the isospin amplitudes $A_{0,2}$
\begin{eqnarray}
A_0 &=&\sum_i [ V_{ud}^*V_{us}c^u_i +V_{cd}^*V_{cs}c^c_i +V_{td}^*V_{ts}c^t_i]
x_i^0 a_0 e^{i\delta^0_s}\;,\nonumber\\
A_2&=&\sum_i [ V_{ud}^*V_{us}c^u_i +V_{cd}^*V_{cs}c^c_i +V_{td}^*V_{ts}c^t_i]
x^2_i a_2 e^{i\delta^2_s}\;.
\end{eqnarray}
 Since in general $x_i^0$ is not equal or proportional to $x^2_i$, $A_0$ 
and $A_2$ do not
have the same weak phases 
\begin{eqnarray}
\delta^w_0 &=& Arg(\sum_i [ V_{ud}^*V_{us}c^u_i +V_{cd}^*V_{cs}c^c_i 
+V_{td}^*V_{ts}c^t_i] x^0_i)\;,\nonumber\\
\delta^w_{2} &=& Arg(\sum_i [ V_{ud}^*V_{us}c^u_i +V_{cd}^*V_{cs}c^c_i 
+V_{td}^*V_{ts}c^t_i]x^2_i)\;.
\end{eqnarray}
Thus the decay amplitude for 
$K^0 \rightarrow \pi^+\pi^-$ can be
parametrized in the form in eq.(\ref{iso}).  The discussion can be 
easily generalized to hyperon decays.

For $K^-\rightarrow \pi^-\pi^0$, the final state has only $I=2$ amplitude,
the decay amplitude is of the form,
\begin{eqnarray}
\tilde A_2&=&\sum_i [ V_{ud}^*V_{us}c^u_i +V_{cd}^*V_{cs}c^c_i 
+V_{td}^*V_{ts}c^t_i]
\tilde x^2_i \tilde a_2 e^{i\delta^2_s}\;.
\end{eqnarray}
It is clear that the particle-antiparticle rate asymmetry vanishes.

For $B\rightarrow \pi\pi$ the situation is, however, very different. We 
will use the same notation for the effective Hamiltonian. Of course
we should keep in mind that now the operators $O_i$ contain a b quark.
Now because the b quark is heavier than a $u$, $\bar u$ pair,
 and also a $c$, $\bar c$ pair, both $c_i^u$, $c_i^c$ have strong rescattering phases 
at the one loop level. At the hadron level, the part responsible for absorptive
amplitude is given by
\begin{eqnarray}
A(\bar B^0\rightarrow \pi^+\pi^-) &=& \sum_i\sum_I<\pi^+\pi^-|[V_{ud}^*V_{ub} c^u_i +  
V_{cd}^*V_{cb} c^c_i +  V_{td}^*V_{tb} c^t_i]|I><I|O_i|\bar B^0>\nonumber\\
&+& \sum_i \sum_{I_c}<\pi^+\pi^-| V_{cd}^*V_{cb} c^c_i
|I_c><I_c|O_i|\bar B^0>\;,
\end{eqnarray}
where I is summed over non-charmed on-shell intermediate states like,
$\pi^+\pi^-$, $\pi^0\pi^0$, ..., and $I_c$ is summed over charmed 
on-shell particle intermediate states like, $D\bar D$ etc.
In this case we can remove the phases in $c_i^u$ because in this decay there
is a tree amplitude with the same CKM factor. However, now
there is another class of  phase shift due to
$<\pi^+\pi^-|c^c_i|I_c>$ which can not be removed. The phase shift of this type to the
lowest order is due to the absorptive part in $c_i^c$.
Because these new phases are generated by
charmed particles in the intermediate state, these phases will 
only appear in the term proportional to $V_{cb}V_{cd}^*$. 
We obtain the $I = 0,\;2$ decay amplitudes $A'_{0,2}$ for $B\rightarrow \pi\pi$
\begin{eqnarray}
A'_0 &=&\sum_i [ V_{ud}^*V_{ub}c^u_i +V_{cd}^*V_{cb}c^c_i +V_{td}^*V_{tb}c^t_i]
x'^0_i a_0 e^{i\delta^{'0}_{s}}\;,\nonumber\\
A'_2&=&\sum_i [ V_{ud}^*V_{ub}c^u_i +V_{cd}^*V_{cb}c^c_i +V_{td}^*V_{tb}c^t_i]
x'^2_i a_2 e^{i\delta^{'2}_{s}}\;.
\end{eqnarray}
Now we should treat $c_i^{u,t}$ to be real and only $c_i^c$ to be complex 
(non-zero rescattering phases) up to order $\alpha_s^2$ in the asymmetry.

Let us now compare these amplitudes with the amplitudes for $K\rightarrow \pi\pi$.
First we note that because in the case for $B\rightarrow \pi\pi$ more 
on-shell intermediate states are allowed, i.e. $\pi\pi$, $\pi\pi\pi\pi$ etc, the
rescattering phases $\delta^{'0,2}_{s}$  
 include inelastic channels unlike the elastic phase shifts $\delta_s^{0,2}$. 
Second
we note that for $B\rightarrow \pi\pi$ there are additional strong rescattering
phases due to on-shell charmed intermediate states. 
The strong rescattering phases
in $c_i^c$ are not canceled by any other contributions. This
has a very important consequence that particle-antiparticle rate asymmetry can occur in a single
isospin amplitude. In fact this happens quite often in $B$ decays\cite{other1}.
 In the following we study four different processes representing 
different
types of $B$ decays, $b\rightarrow \psi s$, $b\rightarrow \phi s$, 
$B^-\rightarrow \eta \pi^-$, and $B^-\rightarrow \eta K^-$.

When considering partial rate asymmetry for single isospin final state in $B$
decays, we do not need
to know the values for the overall strong rescattering phases. We, however, need 
to know
the relative strong rescattering phase shifts between amplitudes with different 
CKM factors by calculating various on-shell rescattering processes.
This calculation is very difficult to carry out. However, we believe that the
WC's and the phases calculated at the quark level could be
good indications of the sizes and the signs of the strong rescattering phases by
appealing to duality. The absorptive parts of hadronic processes are given quite 
accurately by considering the corresponding quark loops as in the calculation of R in
$e^+ e^-$ scattering.
In our later calculation, we will use this approximation. 

In the SM the amplitudes for B decays are generated by the following effective 
Hamiltonian:
\begin{eqnarray}
H_{eff}^q &=& {G_F\over \sqrt{2}}[V_{fb}V^*_{fq}(c_1O_{1f}^q + c_2 O_{2f}^q) -
\sum_{i=3}^{10}(V_{ub}V^*_{uq} c_i^u
+V_{cb}V^*_{cq} c_i^c +V_{tb}V^*_{tq} c_i^t) O_i^q] +H.C.\;,
\end{eqnarray}
where the
superscripts $u,\;c,\;t$ indicate the internal quarks, $f$ can be $u$ or 
$c$ quark. $q$ can be $d$ or $s$ quark depending on if the decay is a $\Delta S = 0$
or $\Delta S = -1$ process.
The operators $O_i^q$ are
defined as
\begin{eqnarray}
O_{f1}^q &=& \bar q_\alpha \gamma_\mu Lu_\beta\bar
u_\beta\gamma^\mu Lb_\alpha\;,\;\;\;\;\;\;O_{2f}^q =\bar q
\gamma_\mu L u\bar
u\gamma^\mu L b\;,\nonumber\\
O_{3,5}^q &=&\bar q \gamma_\mu L b
\bar q' \gamma_\mu L(R) q'\;,\;\;\;\;\;\;\;O_{4,6}^q = \bar q_\alpha
\gamma_\mu Lb_\beta
\bar q'_\beta \gamma_\mu L(R) q'_\alpha\;,\\
O_{7,9}^q &=& {3\over 2}\bar q \gamma_\mu L b  e_{q'}\bar q'
\gamma^\mu R(L)q'\;,\;O_{8,10}^q = {3\over 2}\bar q_\alpha
\gamma_\mu L b_\beta
e_{q'}\bar q'_\beta \gamma_\mu R(L) q'_\alpha\;,\nonumber
\end{eqnarray}
where $R(L) = 1 +(-)\gamma_5$, 
and $q'$ is summed over u, d, and s.  $O_1$ are the tree
level and QCD corrected operators. $O_{3-6}$ are the strong gluon induced
penguin operators, and operators $O_{7-10}$ are due to $\gamma$ and Z exchange,
and ``box'' diagrams at loop level. The WC's $c_i^f$ are defined
at the scale of $\mu \approx
m_b$ which have been evaluated to the next-to-leading order in QCD\cite{buras1,dh1}.
We give the non-zero coefficients below for $m_t = 176$ GeV, $\alpha_s(m_Z) = 0.117$,
and $\mu = m_b = 5$ GeV,
\begin{eqnarray}
c_1 &=& -0.307\;,\;\; c_2 = 1.147\;,\;\;
c^t_3 =0.017\;,\;\; c^t_4 =-0.037\;,\;\;
c^t_5 =0.010\;,
 c^t_6 =-0.045\;,\nonumber\\
c^t_7 &=&-1.24\times 10^{-5}\;,\;\; c_8^t = 3.77\times 10^{-4}\;,\;\;
c_9^t =-0.010\;,\;\; c_{10}^t =2.06\times 10^{-3}\;, \nonumber\\
c_{3,5}^{u,c} &=& -c_{4,6}^{u,c}/N = P^c_s/N\;,\;\;
c_{7,9}^{u,c} = P^{u,c}_e\;,\;\; c_{8,10}^{u,c} = 0
\end{eqnarray}
where $N$ is the number of color, $c^t_i$ are the regularization scheme 
independent WC's obtained in Ref.
\cite{dh1}.
The leading contributions to $P^i_{s,e}$ are given by:
 $P^i_s = (\alpha_s/8\pi) c_2 (10/9 +G(m_i,\mu,q^2))$ and
$P^i_e = (\alpha_{em}/9\pi)(N c_1+ c_2) (10/9 + G(m_i,\mu,q^2))$.  
The function
$G(m,\mu,q^2)$ is give by
\begin{eqnarray}
G(m,\mu,q^2) = 4\int^1_0 x(1-x) \mbox{d}x \mbox{ln}{m^2-x(1-x)q^2\over
\mu^2}\;.
\end{eqnarray}
All the above coefficients are obtained up to one loop order in electroweak 
interactions.
When $q^2 > 4m^2$, $G(m,\mu,q^2)$ becomes imaginary. In our calculation, we will
use $m_u = 5$ MeV, $m_d = 10$ MeV, $m_s = 175$ MeV, $m_c = 1.35$ GeV, and
 the averaged value $m_b^2/2$ for $q^2$.

We must be careful in using absorptive parts of the above WC's. We should always
remove phase shift discussed previously, otherwise one would obtain results violating CPT theorem.

To obtain exclusive decay amplitudes, we need to calculate relevant
hadronic
matrix elements.  Since no reliable calculational tool exists for 
two body modes, 
we shall use factorization 
approximation to get an idea of the size of asymmetry $A_{asy}$. 
The numerical numbers obtained should be viewed as an order of magnitude 
estimates.
The important message is that CP violating partial rate asymmetry in some
single isospin channel decays are indeed non-zero and can reach significant 
magnitude.

\noindent
{\bf Partial rate asymmetry in $b\rightarrow \psi s$}  

Since $\psi$ carries no isospin, the final state
is a single isospin state in this decay.  At the hadronic level, this decay includes
$B\rightarrow \psi K$, $B\rightarrow \psi K^*$, and etc. This decay is 
particularly interesting 
because it has a large branching ratio. CP violating partial 
rate
asymmetry in this type of decay was first studied in the early 80's by 
Brown, Pakvasa and 
Tuan\cite{bpt}. 
Let 
us now analyze this asymmetry using factorization approximation. In this 
approximation,
the partial rate asymmetries for $b\rightarrow \psi s$ and $B^-\rightarrow 
\psi K^-$ are the same. We have
\begin{eqnarray}
&A&(b\rightarrow \psi s) ={G_F\over \sqrt{2}}\bar s \gamma_\mu (1-\gamma_5) b
\{<\psi|\bar u \gamma^\mu u|0>V_{ub}V_{us}^*(c_1 + {c_2\over N})\nonumber\\
&+&<\psi|\bar c \gamma^\mu c|0>[V_{cb}V_{cs}^*(c_1 + {c_2\over N})
-\sum_i V_{ib}V_{is}^*(c_3^i+{c_4^i\over N}+ c_5^i+{c_6^i\over N}
+c_7^i+{c_8^i\over N} + c_9^i+{c_{10}^i\over N})]\}\;.
\end{eqnarray}
Here the first term corresponds to an annihilation contribution which is 
usually small.
If this term is neglected, we have
\begin{eqnarray}
A(b\rightarrow \psi s) &=& {G_F\over \sqrt{2}}\bar s \gamma_\mu (1-\gamma_5) b
<\psi|\bar c \gamma^\mu c|0>\nonumber\\
&\times&[V_{ub}V_{us}^*(C -c^u_7-c_9^u) - V_{cb}V_{cs}^*(C+c_1+{c_2\over N}
-c_7^c-c_9^c)]\;;
\end{eqnarray}
where $C = c_3^t+c_5^t+c_7^t+c_9^t+(c_4^t+c_6^t+c_8^t+c_{10}^t)/N$.
Any phase shift in $c_i^c$ should be removed from our previous 
discussions. 
Only the absorptive part in $c_{7,9}^u$ generate effective 
strong rescattering phases.
The interference which cause the partial rate difference in this case is
of order $\alpha_{em}$ instead naively expected $\alpha_s$ because the 
strong
penguin generated absorptive amplitude cancel in $c_{3,5}^u+c_{4,6}^u/N$. 
In our numerical calculations we will use $N = 2$ favored by experimental 
data, and $|V_{us}| = 0.2205$, $|V_{cb}|=0.04$, and $|V_{ub}/V_{cb}| = 0.08$. 
We find that the partial rate asymmetry is less than $10^{-4}$. The
result is shown in Figure 1. 
One can easily obtain the asymmetry for $b\rightarrow 
\psi d$ by scaling the asymmetry by a factor of $|V_{cb}/V_{ub}|^2$. The 
asymmetry in this case is much large but the branching ratio is much smaller.
In Ref.\cite{soares} using absorptive amplitudes generated by rescattering 
color octet state, asymmetry was estimated for $b\rightarrow \psi d$. 
Using the same calculation for $b\rightarrow \psi s$, we find that the partial 
rate asymmetry is about the same order of magnitude obtained here.
This asymmetry is an order of magnitude smaller than that obtained in 
Ref.\cite{bpt,llc}. Note that here we have neglected the annihilation 
contribution estimated in Ref.\cite{bpt,llc}. However, it is found that when 
current knowledge about CKM parameters is used, the caculation of 
Ref.\cite{bpt,llc} also yields an asymmetry below $10^{-4}$. Hence including
the annihilation diagram is not going to change our result significantly.

\noindent
{\bf Partial rate asymmetry in $b\rightarrow \phi s$}
 
The final state is a single isospin state because $\phi$ is isospin singlet. 
This decay is induced by pure penguin interaction. CP violation in this 
process was first 
evaluated in Ref.\cite{dt}. This process is not affected by 
the previously mentioned effect.  The partial rate asymmetry 
is 
generated by interference of  different penguin amplitude which is of order 
$\alpha_s^2$.  We have
\begin{eqnarray}
A(b\rightarrow \phi s) &=& {G_F\over 2\sqrt{2}} \bar s \gamma_\mu(1-\gamma_5)
b<\phi|\bar s\gamma^\mu s|0>\nonumber\\
&\times&[V_{ub}V_{us}^*(C'-2(c_4^u+{c_3^u\over N}) + c_7^u(1+{1\over N})
+c_9^u)\nonumber\\
&+&V_{cb}V_{cs}^*(C'-2(c_4^u+{c_3^c\over N} + c_7^c(1+{1\over N})
+c_9^c)]\;,
\end{eqnarray}
where $C' = 2(c_3^t+c_4^t+c_5^t)+2(c_3^t+c_4^t+c_6^t)/N -(c_7^t+c_9^t+c_{10}^t
+(c_7^t+c_8^t+c_{10}^t)/N)$. In this case we find the partial rate asymmetry
is of order $O(10^{-3})$. The result is shown in Figure 2.

\noindent 
{\bf Partial rate asymmetry in $B^- \rightarrow \eta \pi^-$}

This is an exclusive decay. Because $\eta$ is an $I = 0$ particle, 
the final state is a single isospin state with $I = 1$. In the factorization 
approximation, we obtain
\begin{eqnarray}
A(B^-\rightarrow \eta \pi^-) &=& {G_F\over \sqrt{2}}
[V_{ub}V_{ud}^*(T_{B\pi}^\eta C^{ut} + T_{B\eta}^\pi D^{ut})\nonumber\\
&+&V_{cb}V_{cd}^*(T_{B\pi}^\eta C^{ct} + T_{B\eta}^\pi D^{ct})]\;,
\end{eqnarray}
where 
\begin{eqnarray}
C^{ut} &=& c_1 + {c_2\over N} - {c_3^u\over 3}-c_4^u - 
+{3\over 2}(c_7^u+{c_8^u\over N})+ {1\over 2}({c_9^u\over N}+c_{10}^u)\nonumber\\
&+&X_d(-2{c_5^u\over N}-2c_6^u+c_7^u+{c_8^u\over N}-c_9^u-{c_{10}^u\over N})
 +\{ c_i^u \rightarrow -c_i^t \}\;,\nonumber\\
D^{ut} &=& {c_1\over N}+c_2 -({c_3^u\over N} +c_4^u +{c_9^u\over N}+
c_{10}^u + 2X_\pi({c_5^u\over N}+ c_6^u+{c_7^u\over N} +c_8^u))
+\{c_i^u \rightarrow -c_i^t\}\;,
\end{eqnarray}
where $X_d = m_\eta^2/(2md(m_b-m_d))$ and $X_\pi = m_\pi^2/(m_u+m_d)(m_b-m_d)$.
$C^{ct}$ and $D^{ct}$ are obtained by setting $c_{1,2}$ to be zero and 
replacing the superscript $u$ by $c$. $T_{B\pi}^\eta$ and $T_{B\eta}^\pi$ 
are defined as,
\begin{eqnarray}
T_{B\pi}^\eta &=& <\pi^-|\bar d\gamma_\mu(1-\gamma_5)b|B^->
<\eta|\bar u\gamma^\mu(1-\gamma_5) u|0>\nonumber\\
&=&i(f_{\eta}/\sqrt{3})F_0^{B\pi}(m^2_\eta)(m_B^2-m_\pi^2)\;,\nonumber\\
T_{B\eta}^\pi &=& <\eta|\bar u \gamma_\mu(1-\gamma_5) b|B^-
<\pi^-\bar d\gamma^\mu(1-\gamma_5) u|0>\nonumber\\
&=&i(f_{\pi}/\sqrt{3})F_0^{B\eta}(m^2_\pi)(m_B^2-m_\eta^2)\;,
\end{eqnarray}
where
$F_0^{Bp}(q^2)$ are the transition form factors between $B$ meson and $p$ meson
(where p could be $\pi$ or $\eta$)
defined in Ref.\cite{gatto,baur}, $f_\pi = 93$MeV, and we will use
$f_\pi \approx f_\eta$. This time any strong phase in $c_i^u$ must be removed.
The interference causing partial rate difference is of order
$\alpha_s$.  The result is shown in Figure 3. 
In the numerical calculation, we have included the $\eta-\eta'$ mixing effect
with the mixing angle $\theta = -20^0$. The figures are ploted with 
$f_{\eta_1}=f_{\eta_8}=f_\pi$. The results are not sensitive the the mixing effect.
The asymmetry can be quite large. 
The branching ratio is 
3 to 4 times larger ($O(10^{-5})$) if the form factors in 
Ref.\cite{gatto} are used than the one obtained using the form factors 
in Ref.\cite{baur}.

\noindent
{\bf Partial rate asymmetry in $B^-\rightarrow \eta K^-$}

The final state is a pure $I = 1/2$ state. We have
\begin{eqnarray}
A(B^-\rightarrow \eta K^-) &=& {G_F\over \sqrt{2}}
[V_{ub}V_{us}^*(T_{BK}^\eta \tilde C^{ut} + T_{B\eta}^K \tilde D^{ut})\nonumber\\
&+&V_{cb}V_{cs}^*(T_{BK}^\eta \tilde C^{ct} + T_{B\eta}^K \tilde D^{ct})]\;,
\nonumber\\
\tilde C^{ut}&=&c_1 +{c_2\over N}+(2{c_3^u\over N} +2c_4^u-{c_9^u\over N}
-c_{10}^u)+{3\over 2}(c_7^u+
{c_8^u\over N} - c_9^u-{c_{10}^u\over N})\nonumber\\
&+&2X_s({c_5^u\over N} +c_6^u - {c_7^u\over N}-c_8^u) + \{c_i^u\rightarrow - 
c_i^t\}
\nonumber\\
\tilde D^{ut}&=&{c_1\over N}+c_2 -{c_3^u\over N}-c_4^u 
-2X_K({c_5^u\over N}+c_6^u+{c_7^u\over N}+c_8^u)\nonumber\\
&-&{c_9^u\over N}-c_{10}^u
+\{c_i^u\rightarrow -c_i^t\}\;,
\end{eqnarray}
where $X_K = m_K^2/(m_s+m_u)(m_b-m_u)$, $X_s = m_\eta^2/(2m_s(m_b-m_s))$, 
$T_{BK}^\eta = i(f_{\eta}/\sqrt{3})F_0^{BK}(m^2_\eta)(m_B^2-m_K^2)$,
$T_{B\eta}^K = i(f_{K}/\sqrt{3})F_0^{B\eta}(m^2_K)(m_B^2-m_\eta^2)$. Similarly the
coefficients $\tilde C_{ct}$ and $\tilde D_{ct}$ are obtained by setting $c_{1,2}$
to be zero, and replacing the superscript $u$ by $c$ in $C_{ut}$ and $D_{ut}$. 
This time the the interference term causing partial rate difference is, again, 
of order
$\alpha_s$. The result is shown in Figure 4. In this case the results are
sensitive to the mixing effect. Within the allowed ranges for $f_{\pi,
\eta_1, \eta_8}$, the branching ratio can change by a factor of 5. The branching ratio
can be as large as $4\times 10^{-6}$ using the form factors in Ref.\cite{gatto}, 
and it is
smaller by a factor of 3 to 4 using  the form factors in Ref. \cite{baur}.
The mixing effect on the asymmetry is less sensitive. 
Again, the asymmetry can be quite large.
For the same values of form factors, we agree with the results obtained 
by Du and Guo in Ref.\cite{other1}. If the electroweak penguin effect is 
neglected, we also agree with Kramer, Palmer and Simma in Ref.\cite{other1}
when the same form factors are used.

To conclude, we have shown that partial rate asymmetry in decays 
of particles containing a strange quark
with single isospin final states, is always zero because the 
allowed intermediate on-shell states are limited. However, the 
situation in B decays is dramatically different. In the latter 
case, more intermediate on-shell states with different CKM factors
are allowed, CP violating partial rate asymmetries need not to 
be zero even if the final state contains only a single isospin state.  

We have carried out detailed analyses for four types of B decays.
 The partial rate asymmetry in $b\rightarrow \psi s$ is small because
the interference causing rate deference in particle and 
anti-particle decay rates is of order $\alpha_{em}$ due to cancellation. 
The partial rate asymmetry in $b\rightarrow \phi s$ is also small 
($O(10^{-3}$). 
The asymmetry in exclusive decays considered here are larger. The
partial rate asymmetry in
$B^-\rightarrow \eta \pi^-$  and  $B^-\rightarrow \eta K^-$ can be quite large.

This work was 
supported in part by the Department of Energy Grant No.
DE-FG06-85ER40224 and DE-AN03-76SF00235. 
XGH is supported by Australian Research Council. XGH would like to thank
Dr. R. Pisarski and the Theory Group of the Brookhaven National Laboratory 
for hospitality where
part of this work was performed.

\begin{figure}[htb]
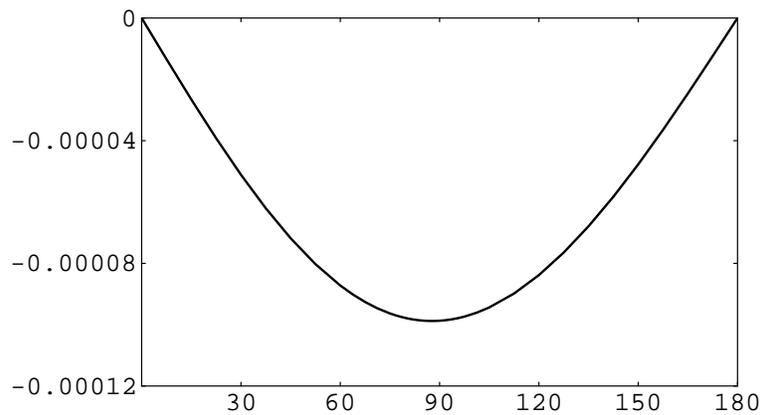

\centerline{ \DESepsf(isospin1.epsf width 10 cm) }
\smallskip
\caption {The partial rate asymmetry for $b\rightarrow \psi s$. 
The vertical axis is the asymmetry and the horizontal axis 
is the value in degree for the phase angle
$\gamma$ in the Wolfenstein parametrization.}
\label{gamma}
\end{figure}

\begin{figure}[htb]
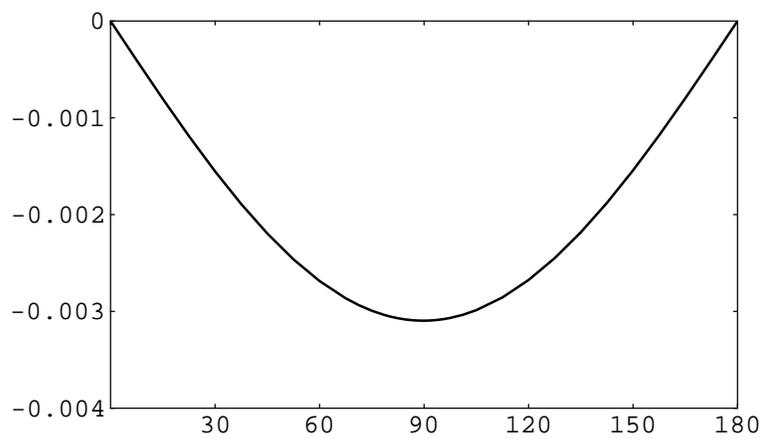

\centerline{ \DESepsf(isospin2.epsf width 10 cm) }
\smallskip
\caption {The partial rate asymmetry for $b\rightarrow \phi s$.
}
\label{gamma1}
\end{figure}

\begin{figure}[htb]
\centerline{ \DESepsf(isospin3.epsf width 10 cm) }
\smallskip
\caption {The partial rate asymmetry for $B^-\rightarrow \eta \pi^-$.}
\label{gamma2}
\end{figure}

\begin{figure}[htb]
\centerline{ \DESepsf(isospin4.epsf width 10 cm) }
\smallskip
\caption {The partial rate asymmetry for $B^-\rightarrow \eta K^-$.}
\label{gamma3}
\end{figure}

\end{document}